\documentclass{PoS}

\newcommand{\gev}{\,\textrm{GeV}}

\newcommand{\eV}{\,\mathrm{eV}}
\newcommand{\Mp}{M_{\rm P}}
\newcommand{\Mpt}{$M_{\rm P}$}
\newcommand{\Mg}{{M_{\rm GUT}}}
\newcommand{\Mgt}{$M_{\rm GUT}$}

\newcommand{\fde}{f_{\rm DE}}
\newcommand{\Ude}{U(1)$_{\rm de}$}
\newcommand{\UPQ}{U(1)$_{\rm PQ}$}

\newcommand{\ndw}{N_{\rm DW}}

\newcommand{\ie}{{\it i.e.~}}
\newcommand{\etal}{{\it et al.}\,}

\title{Axions, strong and weak CP, and  KNP inflation}

\ShortTitle{Axions, strong and weak CP, Inflation}

\author{\speaker{Jihn E. Kim}\thanks{Supported by the NRF grant(No. 2005-0093841) funded by the Korean Government.}\\
          Department of Physics and Astronomy, Seoul National University, 1 Gwanakdaero, Gwanak-Gu, Seoul 151-747, Republic of   
          Korea, and \\
       Department of Physics, Kyung Hee University, 26 Gyungheedaero, Dongdaemun-Gu, Seoul 02447, Republic of Korea\\
        E-mail: \email{jihnekim@gmail.com}}
 
\abstract{ 
I review the ideas leading to the QCD axion and also comment on the Jarlskog determinant describing the  observed weak CP violation, and the axion-related Kim-Nilles-Peloso inflation.  All of these use pseudoscalars, and
the underlying principle is the discrete gauge symmetry either in the bottom-up or top-down approaches. Here, the effects of gravity are required to be unimportant in the low energy effective theory. String compactification is safe from the gravity spoil of global symmetries and some examples from string compactification are commented.  
}

\FullConference{Proceedings of the Corfu Summer Institute 2014 "School and Workshops on Elementary Particle Physics and Gravity",\\
		3-21 September 2014\\
		Corfu, Greece }

\begin{document}

\section{Introduction}

It seems that the most urgent and provable issue in cosmology now is on the nature of the cold dark matter (CDM) which constitutes  27\% in the current cosmic energy pie \cite{PlanckCoParameters}. There is even more dorminant component which is dark energy (DE: currently unknown on its nature), constituting  68\% of the pie. The rest is atoms, neutrinos, etc. Among these, some of DE and CDM can be bosonic coherent motions (BCMs) \cite{BCMrev}. The ongoing searches of CDM are on the weakly interacting massive particles (WIMP) and the QCD axion \cite{Baer15prp}. The QCD axion in cosmology 
\cite{Kim79, KSVZ,DFSZ} is based on the BCM principle \cite{Preskill83}. 
Being a pseudo-Goldstone boson, the QCD axion can be a composite one \cite{KimAxComp}, but after the discovery of the fundamental Brout-Englert-Higgs (BEH) boson, the possibility of the QCD axion being a fundamental particle gained much more weight. Thus, we focus on the possibility of the  fundamental QCD axion.

The QCD axion is very important in two aspect, first it can be a candidate of CDM and second it may be detectable. Since the BEH boson is a fundamental particle, in the bottom-up approach or in an effective field theory framework one can  consider a BEH portal to the high energy scale: to the axion scale or even to the grand unification (GUT) scale. Portals through the BEH boson mainly use the standard model singlets which may be present above the electroweak scale \cite{Kim79}.

Can these singlets explain both DE and CDM in the Universe? Because the axion decay constant $f_a$ can be in the intermediate scale, axions
can live up to now (if $m_a <24\, \eV$) and constitute DM of the
Universe. 

\section{QCD axion}
\label{sec:axion}

\subsection{Pseudoscalar boson}
For the pseudoscalar Goldstone boson, it can be represented as a phase field of the complex spin-0 field $\Phi$,
\begin{equation}
 \Phi=\frac{f+\rho}{\sqrt2} \,e^{i\,a/f},\label{eq:psGoldstone}
\end{equation}
where $\langle\Phi\rangle=f/\sqrt2$ and $\rho$ is the radial field with  $\langle\rho\rangle=0$.
Definition of the pseudoscalar Goldstone boson by (\ref{eq:psGoldstone}) makes sense since it accompanies the U(1) symmetry breaking scale $f$. Then, the leading coupling of $a$ to $\psi$ is proportional to 
 \begin{equation}
 \overline{\psi}_R\psi_L\,fe^{i\,a/f}+{\rm h.c.}=\overline{\psi}i\gamma_5\psi\, a,\label{eq:psCoup}
\end{equation}
which is the desired coupling. The phase field  corresponds to the rotation angle $\alpha$ of the global U(1) symmetry: $\Psi\to e^{i\alpha\gamma_5}\Psi$. A scalar Goldstone boson $s$ cannot be a phase field. The way a scalar Goldstone boson is realized is nonlinear,  
\begin{equation}
s\to s+{\rm constant},\label{eq:scalarshift}
\end{equation}
which can be exponentiated as
\begin{equation}
S=\Lambda\,e^{s/f_s}. 
\end{equation}
Thus, the shift (\ref{eq:scalarshift}) actually changes the scale $\Lambda$, and the scalar $s$ is `dilaton' or  `scale-Goldstone boson'. The symmetry (\ref{eq:scalarshift}) is `dilatonic' symmetry or `scale' symmetry.  
 
For a pseudoscalar Goldstone boson, therefore, we can study the explicit breaking terms more concretely in the top-down approach \cite{KimPLB13}. For simplicity, we consider only one complex field $\Phi$, carrying the U(1) charge,
\begin{equation}
V_{\rm viol}=\sum_n^\infty \frac{c_n}{2\Lambda^{n-4}\,}\Phi^n +{\rm h.c.}=\sum_n^\infty  \frac{|c_n|}{\Lambda^{n-4}\,}\cos\left(\frac{na}{f}+ \delta_n\right),\label{eq:Vviolation}
\end{equation}
where $c_n=|c_n|e^{i\delta_n}$ and $\Lambda$ is a cutoff scale. We require $n$ start from 5 so that the breaking term is small at the scale a high energy scale $\Lambda$. Because of this periodic form,\footnote{
On the other hand, a scalar Goldstone boson does not have this periodic potential.} higher order terms can satisfy some bosonic collective motion (BCM) condition that momenta of each boson are negligible compared to their energy at the onset time of collective oscillation \cite{BCMrev}.  

\subsection{QCD axions} \label{sec:Axions}

Note that the $V-A$ theory was the beginning to the modern particle theory because it reduced the 34 coupling cosntants in Fermi's $\beta$-decay Hamiltonian to just one coupling constant  $G_{\rm F}$. The same kind of reduction on the number of couplings in Eq.  (\ref{eq:Vviolation}) happens if the global symmetry is of the type  `Peccei-Quinn (PQ) symmetry' \cite{PQ77}, \UPQ.
   It is when the global U(1) is broken by a non-Abelian gauge group anomaly, \ie there exists the U(1)$_{\rm global}\times G^2$ where $G$ is a non-Abelian gauge group. Only one U(1)$_{\rm global}$ symmetry breaking term is possible here because the Adler-Bell-Jackiw anomaly \cite{ABJ,AB69} arises only at one-loop, and hence there is only one U(1)$_{\rm global}$ breaking term. This explicit breaking of U(1)$_{\rm global}$ by non-Abelian anomaly was applied to the strong CP problem  by the QCD anomaly. Below the PQ symmetry breaking scale, the resulting Goldstone boson is called {\it axion}\index{axion} and $f$ is called the {\it axion decay constant}\index{axion!axion decay constant} $f_a$.
   
The first thing on which axion is based is the effective interaction in the $\theta$ vacuum \cite{CDG76},
\begin{equation}
{\cal L}=-\frac{\bar{\theta}}{32\pi^2} \,F_{\mu\nu}^a\tilde{F}^{a\,\mu\nu}\label{eq:Lefftheta}
\end{equation}   
where $\bar{\theta}=\theta+{\rm arg.(Det.\,}M_q)$ and $\tilde{F}^{a\,\mu\nu}=\frac12\epsilon^{\mu\nu\rho\sigma}F^a_{
\rho\sigma}$. The coupling $\bar{\theta}$ is a physical one because the interaction (\ref{eq:Lefftheta}) solves \cite{Hooft86} the U(1) problem of QCD \cite{WeinbergU1}. However, the physical $\bar{\theta}$ leads to the so-called {\it strong CP problem}.

The strong CP problem starts from the observed upper bound on the neutron electric dipole moment (NEDM), $d_n$. If a theory of strong interactions violates CP symmetry in full strength, a natural value of NEDM  is expected to be (charge)$\times$(radius) of neutron, \ie O($10^{-13}\,e\,$cm). But, the current lower bound on NEDM is  \cite{NEDMexp},
\begin{equation}
|d_n|< 2.9 \times 10^{-26}\, e\,{\rm cm~ (90\%\, CL)}.\label{eq:NEDMexp}
\end{equation}
First, suppose that  Eq. (\ref{eq:Vviolation}) is the only term violating CP at the strong interaction scale. Then, Eq. (\ref{eq:Vviolation}) implies that the coefficient of CP violating term at the strong interaction scale is less than $\lesssim 10^{-13}$. Thus, we face a fine-tuning problem, setting that coefficient to $\lesssim 10^{-13}$. But with the interaction (\ref{eq:Lefftheta}), we expect \cite{Pospelov05,KimRMP} 
\begin{eqnarray}
d_n(|\bar{\theta}|)\simeq\left\{\begin{array}{l} |\bar{\theta}|\frac{m_*}{\Lambda_{\rm QCD}}\frac{e}{m_n} ~[{\rm with}~m_*=\frac{m_u m_d}{m_u+m_d}]\approx |\bar{\theta}|\cdot(6\times 10^{-16})\, e\,{\rm cm}\\[0.6em]
|\bar{\theta}|\,\frac{g_{\pi n n}}{12\pi^2}\frac{e}{m_n} \ln\left( \frac{m_n}{m_{\pi^0}} \right) \approx |\bar{\theta}|\cdot(4.5\times 10^{-15})\, e\,{\rm cm}
\end{array}
\right.\label{eq:thetabound}
\end{eqnarray}
from which we have $|\bar{\theta}|<10^{-10}-10^{-11}$.

\begin{figure}[!t]
\centerline{\includegraphics[width=0.7\textwidth]{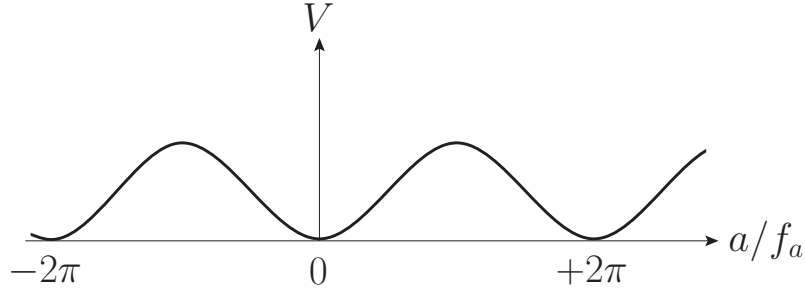} }
\caption{The axion potential with minima at $\bar{\theta}=2\pi n, (n={\rm integer})$. }\label{fig:AxVacuum}
\end{figure}

\begin{figure}[!t]
\centerline{\includegraphics[width=0.25\textwidth]{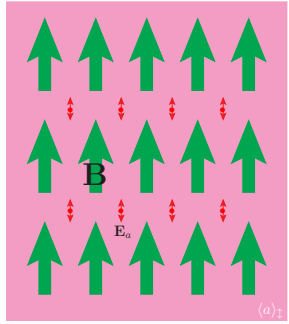} }
\caption{The resonant detection idea of the QCD axion. The {\bf E}-field follows the axion vacuum oscillation.}\label{Fig:AxExp}
\end{figure}

The modern theory of strong interaction QCD is the SU(3) gauge theory with the light quark masses as inputs at the QCD scale around 1\,GeV,
\begin{equation}
{\cal L}_{\rm mass} = -m_u \bar{u}_R u_L-m_d \bar{d}_R d_L-m_s \bar{s}_R s_L+{\rm h.c.}.
\label{eq:quarkmass}
 \end{equation} 
Since the lightest quark $u$ has nonzero mass near 
2.5\,MeV, there is no quark with mass zero \cite{PDG14}. This means that the global symmetry \UPQ~ is explicitly broken by ${\cal L}_{\rm mass} $ of ($\ref{eq:quarkmass}$). In this case, the solution of the strong CP problem is based on the axion potential,
\begin{equation}
V_{\rm axion}\simeq  f_{\pi^0}^2m_{\pi^0}^2 \frac{Z}{(1+Z)^2}\left(1-\cos\frac{a}{f_a} \right),
\end{equation}
where $Z=m_u/m_d$. The potential is shown in Fig. \ref{fig:AxVacuum}. It gives the axion mass
\begin{equation}
m_a=\frac{\sqrt{Z}}{1+Z}\frac{f_{\pi^0}^2m_{\pi^0}^2}{f_a^2}\simeq 0.61\,[\eV]\times\frac{10^7\,\gev}{f_a}.
\end{equation}
The on-going and future axion detection experiments \cite{Yannis15} use cavities in which photons converted from vacuum axions through the Primakoff process are collected in high quality cavity detectors \cite{Sikivie83}. In the cavity, the collective motion of axions in the potential triggers ${\bf E\cdot B}$ oscillating. With the constant {\bf B} field  in the experimental set up,  {\bf E} field oscillates, which is schematically shown in Fig. \ref{Fig:AxExp}. The detection rate has been accurately calculated recently \cite{Hong14}.

\section{Discrete symmetry: mother of all global symmetries}
\begin{figure}[!t]
\centerline{\includegraphics[width=0.25\textwidth]{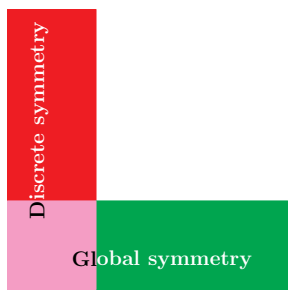}
 }
\caption{Terms respecting discrete and global symmetries.}\label{Fig:discrete}
\end{figure}

For pseudo-Goldstone bosons like axion, we introduce global symmetries. 
 But global symmetries are known to be broken by the quantum gravity effects, especially via the Planck scale  wormholes.   To resolve this dilemma, we can think of two possibilities of discrete symmetries below \Mpt\,\cite{KimPRL13,Kim13worm}: (i) The discrete symmetry arises as  a part  of  a gauge symmetry, 
and (ii) The string selection rules directly give the discrete symmetry. So, we will consider discrete gauge symmetries allowed in string compactification. Even though the Goldstone boson directions in spontaneously broken gauge symmetries are flat, the Goldstone boson directions of spontaneously broken {\em global} symmetries are not flat, \ie global symmetries are always {\em approximate}. The question is what is the degree of the {\em approximateness}. In Fig. \ref{Fig:discrete}, we present a cartoon separating effective terms according to string-allowed discrete symmetries. The terms in the  vertical column represent exact symmmetries such as gauge symmetries and string allowed discrete symmetries. If we consider a few terms in the lavender part, we can consider a {\em global symmetry}. With the global symmetry, we can consider the global symmetric terms which are in the lavender and green parts of Fig. \ref{Fig:discrete}. The global symmetry is broken by the terms in the red part.
The most studied global symmetry is the PQ symmetry \UPQ\,\cite{PQ77} and its physical application to the QCD axion  in terms of the KSVZ axion \cite{KSVZ} and the DFSZ axion \cite{DFSZ}. There are several possibilities even for these: one heavy quark or one pair of BEH doublets  \cite{Kim98}.

\begin{figure}[!t]
\centerline{\includegraphics[width=0.8\textwidth]{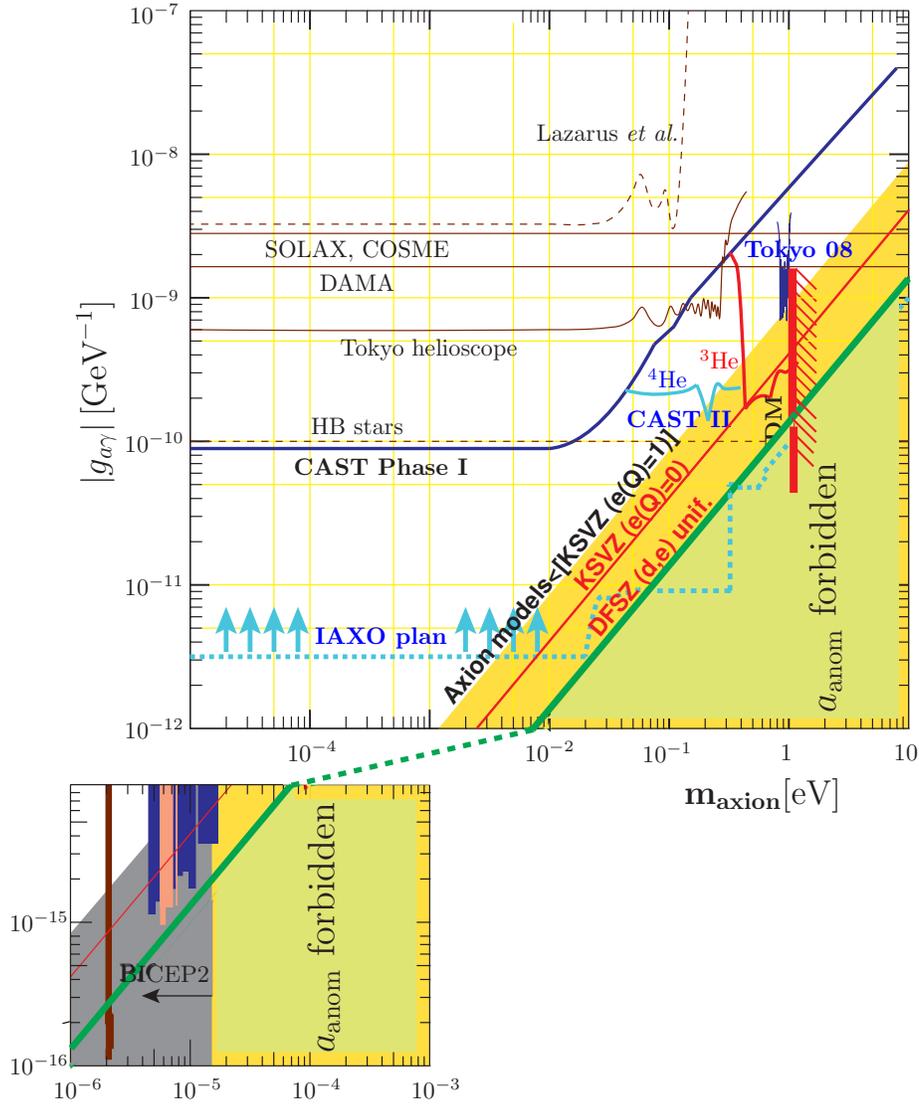} }
\caption{The $g_{a\gamma}(=1.57\times 10^{-10\,}c_{a\gamma\gamma})\,$ vs. $m_a$ plot \cite{BCMrev}.}\label{Fig:expbound}
\end{figure}

For the axion detection through the idea of Fig. \ref{Fig:AxExp}, the axion-photon-photon coupling  $c_{a\gamma\gamma}$ is the key parameter. Here, the afore mentioned gravity spoil of the PQ symmetry applies as pointed out in Refs. \cite{GravSpoil92}.  In this sense, most numbers on $c_{a\gamma\gamma}$ presented in Ref. \cite{Kim98} are ad hoc.  In our search of an ultra-violet completed theory from string, so far there is only one calculation on $c_{a\gamma\gamma}$ calculated in a flipped SU(5) model \cite{Kimagg14}. To calculate $c_{a\gamma\gamma}$, the model must lead to an acceptable SM phenomenology, otherwise the calculation does not lead to a useful global fit to all experimental data. If there is no funnily charged quark and leptons, string theory gives $\bar{c}_{a\gamma\gamma}=\frac83$ \cite{Kimagg14}, which seems to be the minimal one. The proposed axion search experiment in Korea aims at detecting it even if it contributes to CDM of the Universe only  at the level of 10\% \cite{cappsite}. The current search limit is shown in Fig. \ref{Fig:expbound}.

\section{CKM matrix}

The first physics example for discrete symmetry was the parity P in quantum mechanics. The Laporte rule, selecting P eigenstates in centrosymmetric molecules, was known even before the advent of quantum mechanic  \cite{Laporte25}. It was basically the P conservation in quantum electrodynamics(QED).  This discrete symmetry P was later known to be broken outside QED, \ie in weak interactions. In weak interactions, the discrete symmetry CP is also broken.  To observe what will be the relation of the weak CP violation in gravity, we note
 that  CP can be a discrete gauge symmetry in ten-dimensional supergravity \cite{ChoiKaplan}. Thus, string compactification allows CP invariant theories. Then, the weak CP violation  must arise via  spontaneous CP violation method \cite{LeeTD73} at a high energy scale. With this happenning  in string compactification,  there would not be the gravity spoil of the weak-CP discrete symmetry.  Then, the weak CP phase may be a calculable one such as $0,\frac{\pi}{4}, \frac{\pi}{2}$, etc. \cite{KimPLB11}. But, at low energy the nature of CP violation at the high energy scale cannot be probed. The point in this argument is that non-perturbative quantum gravity effects do not require an introduction of new CP odd parameter, which would break CP explicitly. In addition, we can allow compex Yukawa couplings if it results from string compactification \cite{KimPLB13}.   These may appear as complex Yukawa couplings in the SM from the VEVs of complex structure moduli  and the vacuum angles arising from the VEVs of stringy axions \cite{StromingerWitten}. Whether the weak CP phase is determined spontaneously at the high energy scale or not, we can consider a general weak CP phase at low energy. If there is one pair of Higgs doublets in SUSY SM, the only other complex parameter is the $\mu$ term the  phase of which does not affect the CKM matrix. Therefore, it would be convenient that the physically measured weak CP phase $\delta$ clearly appears in the CKM matrix itself. For this purpose, we presented a new form for the CKM matrix \cite{KimSeo11},
\begin{eqnarray}
 \left(\begin{array}{ccc} c_1,&s_1c_3,&s_1s_3 \\ [0.2em]
 -c_2s_1,&e^{-i\delta}s_2s_3 +c_1c_2c_3,&-e^{-i\delta} s_2c_3+c_1c_2s_3\\[0.2em]
-e^{i\delta} s_1s_2,&-c_2s_3 +c_1s_2c_3 e^{i\delta},& c_2c_3 +c_1s_2s_3 e^{i\delta}
\end{array}\right) \label{eq:KSexact}
\end{eqnarray}
where $c_i=\cos\theta_i$ and $s_i=\sin\theta_i$. We will call the phase $\delta$ appearing in $V_{31}$ the ``Jarlskog invariant phase''  since it is the physical phase describing the strength of the weak CP violation \cite{Jarlskog85}. In this form, the Jarlskog determinant $J$ can be expressed as the magnitude of the imaginary part of $V_{31}V_{22}V_{13}$, which can be proven as follows \cite{KimMo15}. Let the determinant of $V$ be real as for Eq. (\ref{eq:KSexact}).
For the real and hence the unit determinant, ${\rm Det.}\,V =1$, multiply   $V_{13}^*V_{22}^*V_{31}^*$ on both sides. Then, we obain  
\begin{eqnarray}
V_{13}^* &V_{22}^*V_{31}^* =|V_{22}|^2V_{11}V_{33}V_{13}^*V_{31}^*-V_{11}
 V_{23}V_{32}V_{13}^*V_{31}^*V_{22}^* \nonumber\\
 &+|V_{31}|^2V_{12}V_{23}V_{13}^*V_{22}^* -V_{12}V_{21}V_{33}V_{13}^*V_{31}^*V_{22}^*\\
 &+|V_{13}|^2V_{21}V_{32}V_{31}^*V_{22}^*-|V_{13}V_{22}V_{31}|^2.\nonumber
 \end{eqnarray}
Using the unitarity of $V$, this equation can be rewritten as
\begin{eqnarray}
   &&V_{13}^*V_{22}^*V_{31}^* =(1-|V_{21}|^2)V_{11}V_{33}V_{13}^*V_{31}^* 
  +V_{11}V_{23} V_{13}^*V_{21}^*|V_{31}|^2\nonumber\\ 
 && +(1-|V_{11}|^2)V_{12} V_{23}V_{13}^*V_{22}^* +|V_{13}|^2(V_{12}V_{21}V_{11}^*V_{22}^*
  +V_{21}V_{32}V_{31}^*V_{22}^*)
 -|V_{13}V_{22}V_{31}|^2.\label{eq:detmult3}
\end{eqnarray}
Let the imaginary part of $V_{11}V_{33}V_{13}^*V_{31}^*$ be $J$. Now, using the unitarity relations again, we can express the imaginary part of the RHS of Eq. (\ref{eq:detmult3}) as
$[(1-|V_{21}|^2)-|V_{31}|^2+(1-|V_{11}|^2)]J=J$.
Therefore, the imaginary part of  $V_{13}^*V_{22}^*V_{31}^*$ (the  LHS of Eq. (\ref{eq:detmult3})) is $J$.  It is  the imaginary part of any one element among the six components of determinant of $V$, for example  $J=|{\rm Im}\,V_{13} V_{22} V_{31} |$. This simplifies how the weak CP violation is scrutinized just from  the CKM matrix elements. For example, if the elements of the first row is made real then the phase of $V_{31}$ is an invariant phase \cite{KimMo15}. This simple form for $J$ has not been known almost three decades \cite{WeakCPbook}.

\begin{figure}[!t]
  \begin{center}
 \begin{tabular}{c}
 \includegraphics[width=0.27\textwidth]{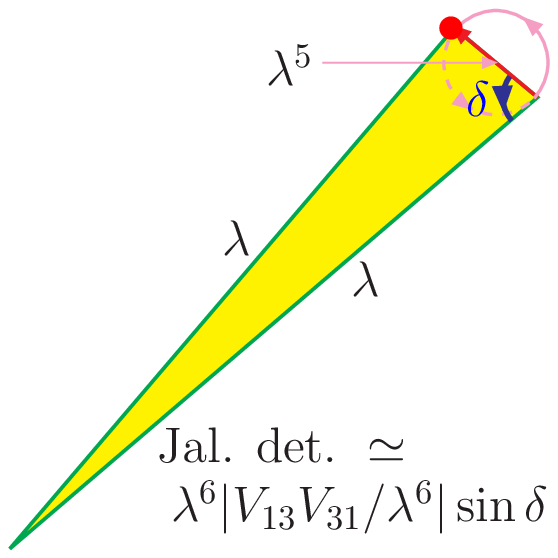} \\ \hskip 4.5cm (a)\\
    \includegraphics[width=0.35\textwidth]{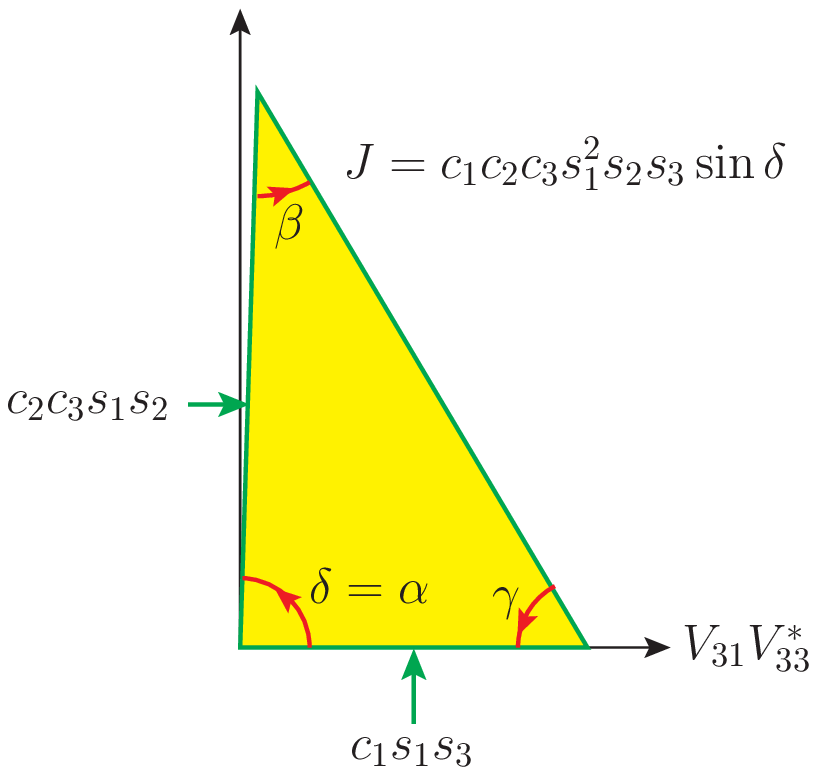} \\ \hskip 4.5cm  (b)
  \end{tabular}
  \end{center}
 \caption{The Jarlskog triangles. (a) The Jarlskog triangle with  two long sides of $O(\lambda)$.  (b) The Jarlskog triangle with the first and third columns. The magnitude of the phase from the almost vertical green line comes from the coefficient of the factor $e^{-i\delta}$ of $V_{13}^*V_{22}^*V_{31}^*$ of Eq. (4.1).}
\label{fig:JTriLong}
\end{figure}

For any Jarlskog triangle, the area is the same. With the $\lambda= \sin\theta_1\equiv \sin\theta_C$ expansion,  $J$ is of order $\lambda^6$.  Figure \ref{fig:JTriLong}\,(a) shows  two long sides. Rotating the $O(\lambda^5)$ side (the red arrow), the CP phase $\delta$ and also the area change and it is maximal with $\delta\simeq\frac{\pi}{2}$. The maximality of $\delta$ is therefore physical. As $\delta$ is rotated in  Fig. \ref{fig:JTriLong}\,(a), the Jarlskog triangle of Fig.  \ref{fig:JTriLong}\,(b) also rotates the shape and its area becomes maximum when $\delta\simeq \pi/2$. From the PDG value of $\delta\simeq 90^{\rm o}$ for the triangle of the type Fig.  \ref{fig:JTriLong}\,(b), we determine $\delta\simeq 90^{\rm o}$ \cite{KimMo15}, which is actually determined as $\alpha$ in Ref. \cite{PDG14}. At present, this simple fact is not known to many weak CP experts.  In Ref. \cite{KimMo15}, using this   $\delta$, the  final state strong-interaction phases are determined from the data on $\overline{B}_{d,s}^0\to K^{-} \pi^+$ \cite{LHCb13}.

\section{Dark energy}
\label{sec:axionicde}

It is interesting to note that the QCD axion must arise if one tries to introduce the DE scale via the idea of Fig. \ref{Fig:discrete} \cite{KimNilles14,KimJKPS14}. The DE and QCD axions are the BCM examples. Dark energy  is classified as {\bf CCtmp} and QCD axion  is classified as {\bf BCM1} in  \cite{BCMrev}. 

We argued that mother of all global symmetries are the discrete symmetry from string compactification \cite{Baer15prp}. This applies to the QCD axion. If we want to interpret the DE scale by a tiny height of DE potential from spontaneously broken \Ude, a {\bf CCtmp} pseudoscalar mass is in the range $10^{-33}\sim 10^{-32}\,\eV$ \cite{Carroll98}. But, the QCD axion mass in the range of milli- to nano-eV for $f_a\simeq 10^{9-15}\,\gev$. Therefore, the QCD anomaly term is too large to account for the DE scale of $10^{-46\,}\gev^4$, and we must find out a QCD-anomaly free global symmetry. It is possible by introducing two global U(1) symmetries  \cite{KimNilles14,KimJKPS14}. 
 For the DE pseudo-Goldstone boson not to have oscillated yet, the breaking scale of \Ude~is trans-Planckian \cite{Carroll98}. 
\begin{figure}[!t]
\centerline{\includegraphics[width=0.5\textwidth]{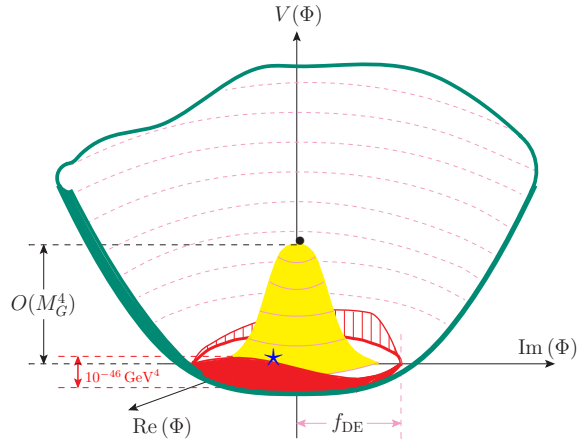}
 }
\caption{The DE potential in the red angle direction in the valley of radial field of height  $\approx \Mg^4$.}\label{Fig:HilltopV}
\end{figure}

 A recent calculation of the cosmic axion density, including the anharmonic term carefully, gives the following axion window \cite{Bae08},
\begin{equation} 
10^{9\,}\gev<f_a<10^{12\,}\gev.\label{eq:window}
\end{equation}
 It has been argued that the QCD axion is better from matter fields  \cite{KimPRL13}.  To interpret the DE scale, introduction of two global symmetries, \ie   \UPQ\,and \Ude,  is inevitable \cite{KimJKPS14}.  Hence, the appearance of \UPQ~is a natural consequence.
 The decay constant corresponding to \Ude, $\fde$,  is so small $10^{-46\,}\gev^4$ that the needed discrete symmetry breaking term of Fig. \ref{Fig:discrete} must be small, \ie the discrete symmetry must be of high order. Then, we have a scheme to explain both 68\% of DE and 27\% of CDM via approximate {\em global} symmetries. With SUSY, axino may contribute to CDM also \cite{CKR00}.
 Using the SUSY language,
the discrete and global symmetries below $\Mp$ are the consequence of the full superpotential $W$. So, the exact symmetries related to string compactification are respected by the full $W$, \ie the vertical column of Fig. \ref{Fig:discrete}. Considering only the $d=3$ superpotential $W_3$,  we can consider an
approximate PQ symmetry. 

A typical example for the  discrete symmetry is ${\bf Z}_{10\,R}$ as shown in \cite{KimJKPS14}. The ${\bf Z}_{10\,R}$  charges descend from a gauge U(1) charges of the string compactification \cite{KimPLB13}.  In this scheme with BEH portal, we introduced three  scales for vacuum expectation values (VEVs), TeV scale for $H_uH_d$, the GUT scale \Mgt~for singlet VEVs, and the intermediate scale for the QCD axion. The other fundamental scale is $\Mp$. The trans-Planckian decay constant $\fde$ can be a derived scale \cite{KNP05}. The TeV scale can be derived from the intermediate scale, or vice versa \cite{KimNilles84}.

The height of the \Ude~potential is  order $\Mg^4$. One consequence of this  potential is a hilltop inflation with the height of $O(\Mg^4)$, as shown in Fig. \ref{Fig:HilltopV}. It belongs to a small field inflation, consistent with the 2013 Planck data.

\section{Gravity waves  from U(1)$_{\rm de}$ potential}
\label{sec:gravitywave}
 
After the talk, however, we were informed of the surprising report from the BICEP2 group on a large tensor-to-scalar ratio $r$  \cite{BICEP2}. But, a later 2015 Planck report \cite{Planck2015} suggests that $r$ may not be as large as the initial BICEP2 report.  Anyway, we must reconsider the above hilltop inflation whether it leads to appropriate numbers on $n_s, r$ and the e-fold number $e$, or not.
 With two U(1)'s, the large trans-Planckian $\fde$ is not spoiled by the intermediate PQ scale $f_a$ because the PQ scale just adds to the $\fde$ decay constant only by a tiny amount, viz. $\fde\to \sqrt{\fde^2+O(1)\times f_a^2} \approx\fde$ for $|f_a /\fde\simeq 10^{-7}|$.  Therefore, for a high scale inflation, we consider only \Ude~without a QCD anomaly.

Any models can lead to inflation if the potential is flat enough as in the chaotic inflation with small parameters \cite{Linde83}. A single field chaotic inflation example for the BICEP2 report is the $m^2\phi^2$ scenario  with $m=O(10^{13\,}\gev)$. To shrink the field energy much lower than $\Mp^4$, a   natural inflation (mimicking the axion-type minus $\cos$ potential) was introduced \cite{Freese90}. If a large $r$ were to be observed, Lyth long time ago noted that the field value $\langle\phi\rangle$ must be trans-Planckian, $\gtrsim 15\,\Mp$,  the so-called Lyth bound \cite{Lyth97}. To obtain the trans-Planckian field value, the Kim-Nilles-Peloso (KNP) 2-flation has been introduced with two axions \cite{KNP05}. It is known recently that the natural inflation is more than $2\sigma$ away from the central value of BICEP2, $(r,n_s)=(0.2,0.96)$. In general, the hilltop inflation  gives almost zero $r$.  
 
For the \Ude\,hilltop inflation to give a suitable $n_s$ with a large $r$, it is necessary to introduce another field so that it provides the behavior of $m^2\phi^2$ term at the BICEP2 point \cite{KimHilltop14}. With this corrected hilltop potential, the height is of order $\Mg^4$ and the decay constant can be $> 15\,\Mp$ \cite{Lyth97}. In the hilltop potential, the potential energy is smaller than order $\Mp^4$  for
$\phi=[0,\fde]$. Since this hilltop potential is obtained from the mother discrete symmetry, such as ${\bf Z}_{10\,R}$, the flat valley up to the trans-Planckian $\fde$ is possible, for which the necessary condition is given in terms of quantum numbers of  ${\bf Z}_{10\,R}$ \cite{KimHilltop14}.
 
\section{The KNP model and \Ude~hilltop inflation}
\label{sec:KNPtrans}
 
Next two sections are added after the talk since inflation ideas after the BICEP2 report, following the axion-type potential, were widely used under the name of {\em natural inflation}.
Inflation needs a region of field space where the potential is almost flat. The first example was the Coleman-Weinberg potential, where logarithmic term can be considered almost flat. In most inflationary models, flat regions are assumed in the inflaton potential. Another example for the flat potential is the axion potential because there are two scales in the potential, the PQ symmetry breaking scale $\Lambda$ and the axion decay constant $f$ which is considered to be much larger than $\Lambda$ as in most axion models. This axion type potential was used in inflation with $\Lambda\simeq\Mg$ and $f\lesssim\Mp$ \cite{Freese90}.

But the natural inflation of Ref. \cite{Freese90} cannot raise the axion decay constant above $\Mp$. To resolve this dilemma, Ref. \cite{KNP05}  introduced two axions with the potential,
\begin{eqnarray}
V = \Lambda_{1}^4 \left(1-\cos\left[\alpha\frac{a_1}{f_1}+\beta\frac{a_2}{f_2}\right]\right)
 +  \Lambda_{2}^4 \left(1-\cos\left[\gamma\frac{a_1}{f_1}+\delta\frac{a_2}{f_2}\right]\right),\label{eq:KNP2}
\end{eqnarray}
where  $f_1$ and $f_2$ are O(\Mgt)  $\alpha,\beta,\gamma$, and $\delta$ are determined by two U(1) PQ quantum numbers,  and $\Lambda_1$ and $\Lambda_2$ are two confining scales of nonabelian gauge groups. 
From this potential, we calculate the mass matrix of two axions
\begin{equation}
M^2=\left(\begin{array}{cc}
   \frac{1}{f_1^2}\left({\alpha^2\Lambda^4_1}+ {\gamma^2\Lambda^4_2}
   \right), &\frac{1}{f_1f_2} (\alpha\beta\Lambda^4_1+ \gamma\delta\Lambda^4_2) 
   \\[1em]
   \frac{1}{f_1f_2}(\alpha\beta\Lambda^4_1+ \gamma\delta\Lambda^4_2) , &   
 \frac{1}{f_2^2}(\beta^2\Lambda^4_1+\delta^2\Lambda^4_2) \end{array}
\right).\nonumber\label{eq:Msquare}
 \end{equation}
Mass eigenvalues are
  \begin{eqnarray} 
 m_{a_h}^2=\frac12(A+B), ~
 m_{a_I}^2=\frac12(A-B),\label{eq:mInf}
 \end{eqnarray} 
 where
  \begin{eqnarray} 
 A =  \frac{\alpha^2\Lambda_1^4 
+\gamma^2\Lambda_2^4}{f_1^2} +\frac{\beta^2\Lambda_1^4 
+\delta^2\Lambda_2^4}{f_2^2}  \label{eq:A},~
B =  \sqrt{A^2  -4(\alpha\delta -\beta\gamma)^2 \frac{\Lambda_1^4 \Lambda_2^4}{f_1^2 f_2^2} }\,. \label{eq:B}
 \end{eqnarray} 
 
\begin{figure}[!t]
\centerline{\includegraphics[width=0.45\textwidth]{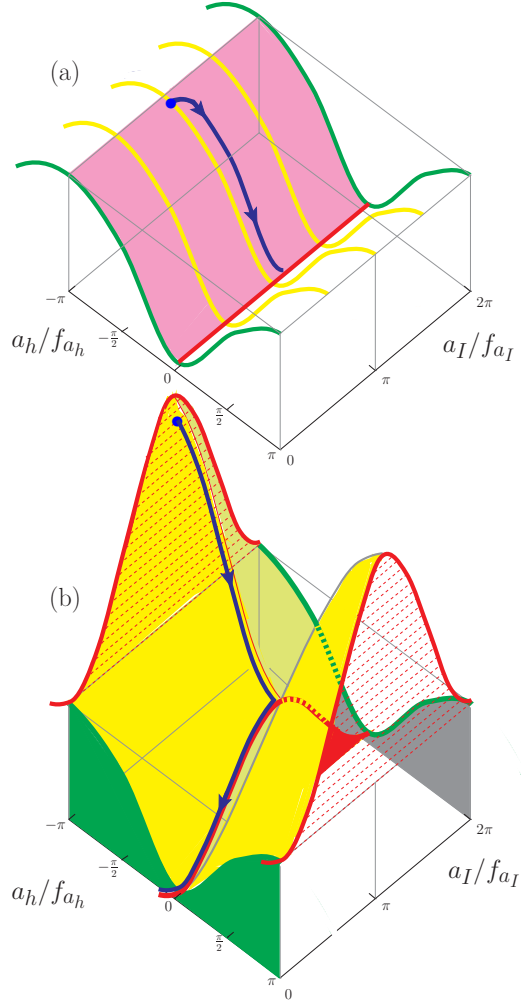}   }\vskip 0.9cm
\caption{Two-flation shown in \cite{CKKyae15}. (a) The flat valley with one confining force, and (b) the KNP model with two confining forces.}\label{Fig:twofl}
\end{figure}

From Eqs. (\ref{eq:B}),  the inflaton mass $m_{a_I}$ is known to vanish for $\alpha\delta=\beta\gamma$, which corresponds to an infinite $f_{a_I}$. Thus, a large $f_{a_I}$ is possible for $\alpha\delta\approx \beta\gamma$. A small number $\Delta$ parametrize this, \ie $\alpha\delta=\beta\gamma+\Delta$.
Then, the heavy axion and  inflaton masses are
\begin{eqnarray}
m_{a_h}^2 \simeq \frac{\alpha^2\Lambda_1^4 
+\gamma^2\Lambda_2^4}{f_1^2} +\frac{\beta^2\Lambda_1^4 
+\delta^2\Lambda_2^4}{f_2^2},~~
m_{a_I}^2  \simeq  \frac{\Delta^2\Lambda_1^4\Lambda_2^4}{D}
  \end{eqnarray}
  where ${D=f_2^2(\alpha^2\Lambda_ 1^4 +\gamma^2\Lambda_2^4) +f_1^2(\beta^2\Lambda_1^4 
+\delta^2\Lambda_2^4)}.$
For simplicity, let us discuss for $\Lambda_1=\Lambda_2=\Lambda$ and $f_1=f_2\equiv f$. Then, the masses are
\begin{eqnarray}
m_{a_h}^2 \simeq (\alpha^2 
+\beta^2 +\gamma^2+\delta^2 )\frac{\Lambda^4}{f^2},~~
 m_{a_I}^2 \simeq  \frac{\Lambda^4}{(\alpha^2 +\beta^2 +\gamma^2+\delta^2 )f^2/\Delta^2}\,,
\label{eq:mIapp} 
 \end{eqnarray}
 from which we obtain \cite{CKKyae15},
 \begin{equation}
 f_{a_I}=\frac{\sqrt{\alpha^2 +\beta^2 +\gamma^2+\delta^2 }f}{|\Delta|}.
 \label{eq:faI} 
 \end{equation}
With the same order of $\alpha,\beta,\gamma,$ and $\delta$, the small number $\Delta$ can be O($1$) to realize $f_{a_I}\approx 100f$ if $\alpha,\beta,\gamma,\delta=$ O(50).  This potential is depicted in Fig. \ref{Fig:twofl} \cite{CKKyae15}.
 
\section{PQ symmetry breaking below $H_I$} \label{sec:PQscale}

The needed axion scale given in Eq. (\ref{eq:window}), far below the GUT scale, is understood in models with the anomalous U(1) in string compactification \cite{Kim88,KimPRL13}. In addition to the scale problem, there exists the cosmic-string and domain wall (DW) problem \cite{Vilenkin82,Sikivie82}. Here, I want to stress that the axion DW problem
has to be resolved without the dilution effect by inflation. 

\begin{figure}[!t]
\centerline{\includegraphics[width=0.4\textwidth]{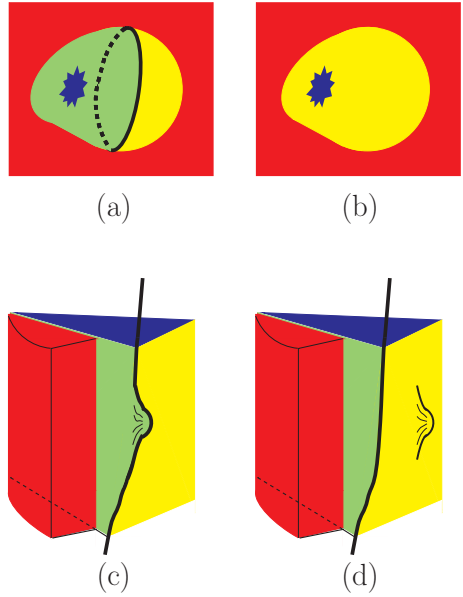} }
\caption{Small DW balls ((a) and (b), with punches showing the inside blue-vacuum) and the horizon scale string-wall system ((c) and (d))  for $\ndw=2$. Yellow walls are $\theta=0$ walls, and  yellow-green walls are $\theta=\pi$ walls.  Yellow-green walls of type (b) are also present.}\label{Fig:DWtwo}
\end{figure}

\begin{figure}[!t]
\centerline{\includegraphics[width=0.5\textwidth]{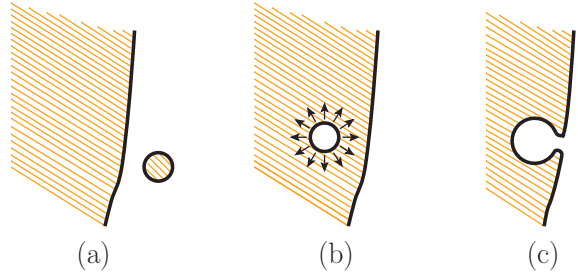} }
\caption{The horizon scale string-wall system with $\ndw=1$. Any point is connected to another point, not passing through the wall [58]. }\label{Fig:DWone}
\end{figure}

The reheating temperature after inflation may be $\gtrsim 10^{12\,}\gev$, which is the case if the BICEP2 finding of $r$ is not a few orders smaller than the initial report. Then, studies on the isocurvature constraint  
pin down the axion mass in the upper allowed region of Fig. \ref{Fig:expbound} \cite{Marsh14}.  But this axion mass is based on the numerical study of Ref. \cite{Kawasaki12} which has not included the effects of axion string-DW annihilation by the Vilenkin-Everett mechanism \cite{Vilenkin82}.   In Fig. \ref{Fig:DWtwo}, we present the case for $\ndw=2$. Topological defects are small balls ((a) and (b)), whose walls separarte $\theta=0$ and $\theta=\pi$ vacua, and a horizon scale string-wall system. Collisions of small balls on the horizon scale walls do not punch a hole, and the horizon size string-DW system is not erased ((c) and (d)). Therefore, for $\ndw\ge 2$ axion models, there exists the cosmic energy crisis problem of the string-DW system.   In Fig. \ref{Fig:DWone},   the case for $\ndw=1$ is presented. Topological defects are small disks and a horizon scale string-DW system ((a)). Collisions of small balls on the horizon scale walls punch holes ((b)), and the holes expand with light velocity. In this way, the string-wall system is erased  ((c)) and the cosmic energy crisis problem is not present in $\ndw=1$ axion models \cite{BarrChoiKim}, for example with one heavy quark in the KSVZ model. If the horizon-scale string-DW system is absent, there is no severe axion DW problem. 

At present with the possibility of a large $r$, it is important to realize $\ndw=1$ axion models. One well-known example is the so-called Lazarides-Shafi mechanism, using the center (discrete group) of GUT gauge groups \cite{LS82}. A more useful discrete group is a discrete subgroup of continuous U(1)'s, \ie the discrete points of the longitudinal Goldstone boson directions of gauged U(1)'s \cite{ChoiKimDW85}. It was pointed out that the anomalous gauged U(1) is useful for this purpose in string theory \cite{Kim88}.  This solution has been recently pointed out in ${\bf Z}_{12-I}$ orbifold compactification \cite{KimDW14}.
 
The QCD-axion string-DW problem may not appear at all if the hidden-sector confining gauge theory conspire to erase the hidden-sector string-DW system \cite{BarrKim14}. Here,  we introduce just one axion, namely through the anomalous U(1) gauge group, surviving down to the axion window as a global \UPQ.  In addition, we introduce two kinds of heavy quarks, one the SU($N_h$) heavy quark $Q_h$ and the other SU(3)$_{\rm QCD}$ heavy quark $q$. 
\begin{figure}[!h]
\centerline{\includegraphics[width=0.38\textwidth]{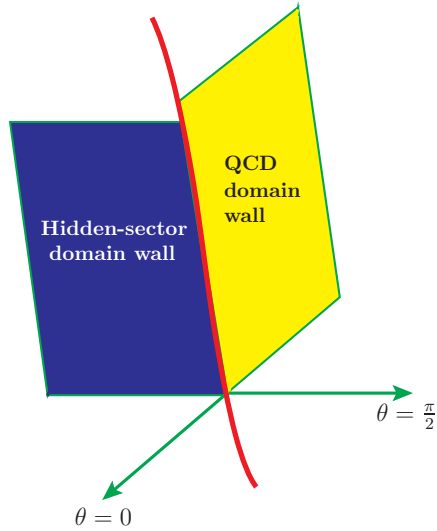} }
\caption{Two walls: the QCD and the hidden sector walls, attached to the axion string. Here, only one PQ symmetry is present. The relative angle $\frac{\pi}{2}$ is just for an illustration. }\label{Fig:DWhidden}
\end{figure}
Then, the type of  Fig. \ref{Fig:DWhidden}  is present with two kinds of walls: one of $\Lambda_{h}$ wall and the other of $\Lambda_{\rm QCD}$ wall. But, at $T\approx \Lambda_{h}$ only $\Lambda_{h}$ wall is attached. At somewhat lower temperature $T_{\rm lower}$ ($< \Lambda_{h}$) the string-DW system is erased {\it \`a la} Fig. \ref{Fig:DWone}. The height of the  $\Lambda_{h}$ wall is proportional to $m_{Q_h}\Lambda_h^3$ with $m_{Q_h}=f\langle X\rangle $. The VEV $\langle X\rangle $ is temperature dependent, and it is possible that  $\langle X\rangle =0$ below some critical temperature $T_c\,(<T_{\rm lower})$. Then, the  $\Lambda_{h}$ wall  is erased below $T_c$, and at the QCD phase transition only the QCD wall is present. But, all horizon scale strings have been erased already and there is no energy crisis problem of the QCD-axion string-DW system. Therefore, pinpointing the axion mass using the numerical study of Ref. \cite{Kawasaki12} is not water-proof.

The $\ndw=1$ models are very attractive and it has been argued that the model-independent axion in string models, surviving down as a \UPQ\,symmetry below the anomalous U(1) gauge boson mass scale, is good for this. At the intermediate mass scale $Q_{\rm PQ}=1$ should obain a VEV to have  $\ndw=1$. In a ${\bf Z}_{12-I}$ orbifold compactification with the anomalous U(1), the axion-photon-photon coupling has been calculated \cite{Kimagg14},
\begin{equation} 
c_{a\gamma\gamma}=
\overline{c}_{a\gamma\gamma}-0.98 \gtrsim \frac83 -0.98\simeq  0.69.
\end{equation}  
With the electroweak hypercharge form $Y={\rm diag.}(-\frac13,-\frac13,-\frac13, \frac12, \frac12,c,\cdots)$ in GUTs, we have $\sin^2\theta_W=3/(8+6[c^2+\cdots])$ \cite{Kim80prl,CKKNRV15} and $\overline{c}_{a\gamma\gamma}\ge \frac83$. In Fig. \ref{Fig:expbound}, the forbidden region from the anomalous U(1) is shown as light green. The lower bound line is the same as the $(d^c,\, e)$ unification DFSZ line. But,  the PQ symmetry being approximate, if some approximate symmetry \cite{IWKim07} is used for the QCD axion, then the region  below the light green line of Fig. \ref{Fig:expbound}  is also allowed.
 

\end{document}